\documentclass[prb,twocolumn,eqsecnum,showpacs]{revtex4}

\usepackage{graphicx}

\begin{document}  

\title{Many-body corrections to cyclotron resonance 
in monolayer and bilayer graphene
}
\author{K. Shizuya}
\affiliation{Yukawa Institute for Theoretical Physics\\
Kyoto University,~Kyoto 606-8502,~Japan }

\begin{abstract} 

Cyclotron resonance in graphene is studied with focus on many-body corrections 
to the resonance energies,
which evade Kohn's theorem.
The genuine many-body corrections turn out to derive from vacuum polarization, 
specific to graphene, which diverges at short wavelengths.
Special emphasis is placed on the need for renormalization, which allows one 
to determine many-body corrections uniquely from one resonance to another. 
For bilayer graphene, in particular, both intralayer and interlayer coupling strengths
undergo infinite renormalization; 
as a result, the renormalized velocity and interlayer coupling strength run 
with the magnetic field. 
A comparison of theory with the experimental data is made 
for both monolayer and bilayer graphene.
 
\end{abstract} 


\maketitle

\section{Introduction} 

Graphene, a monolayer graphite,  attracts great attention 
for its unusual electronic transport~\cite{NG,ZTSK,ZJS,ZA,GS,PGN}
as well as its potential applications.
It supports as charge carriers massless Dirac fermions,
which lead to such exotic phenomena as 
the half-integer quantum Hall effect and minimal conductivity.

The multi-spinor character of the electrons in graphene derives 
from the sublattice structure of the underlying honeycomb lattice, 
  and this immediately implies, in the low-energy effective
theory of graphene, the quantum nature of the vacuum state;~\cite{Semenoff}
the conduction and valence bands are related by charge conjugation  
and, in particular, the latter acts as the Dirac sea.
Graphene in a magnetic field $B$ thus gives rise to 
a particle-hole symmetric $\lq\lq$relativistic" pattern of Landau levels, 
with spectra 
$\epsilon_{n}\propto \pm \sqrt{|n|}\sqrt{B}$ unequally spaced, 
together with four characteristic zero-energy Landau levels 
(whose presence has a topological origin~\cite{NS}). 

This nontrivial vacuum structure is the key feature 
that distinguishes graphene and its multilayers 
from conventional quantum Hall (QH) systems.
In particular, bilayer graphene~\cite{NMMKF} supports, 
as a result of interlayer coupling,   massive fermions, 
which, in a magnetic field, again develop 
a particle-hole symmetric tower of Landau levels,
with an octet of zero-energy levels.~\cite{MF}
Bilayer graphene has a unique property that its band gap 
is externally controllable.~\cite{OBSHR,Mc,CNMPL,OHL}

Graphene and its multilayers give rise to rich spectra 
of cyclotron resonance, with resonance energies varying from one transition 
to another within the electron band or the hole band, and, notably,
even between the two bands.
This is in sharp contrast with conventional QH systems with a parabolic dispersion,
where cyclotron resonance (optically-induced at zero momentum transfer ${\bf k}=0$)
takes place between adjacent Landau levels, hence at a single frequency 
$\omega_{c}= eB/m^{*}$ which, according to Kohn's theorem,~\cite{Kohn} 
is unaffected by Coulomb interactions.
Nonparabolicity~\cite{AA} of the electronic spectra in graphene 
evades Kohn's theorem and 
offers the possibility to detect the many-body corrections to cyclotron resonance,
as discussed theoretically for monolayer graphene.~\cite{IWFB,BMgr}

Experiment has already studied via infrared spectroscopy 
cyclotron resonance in monolayer~\cite{JHT,DCN} 
and bilayer~\cite{HJTS} graphene,
and verified the characteristic 
features of the associated Landau levels.  
Data generally show no clear sign of the many-body effect, 
except for one~\cite{JHT} on monolayer graphene.

The purpose of this paper is to study the many-body effect 
on cyclotron resonance in graphene, 
by constructing an effective theory of cyclotron resonance
within the single-mode approximation.
It is shown that the genuine many-body corrections 
arise from vacuum polarization, specific to graphene, 
which actually diverges logarithmically at short wavelengths
and requires renormalization.
Our approach in part recovers results of earlier studies~\cite{IWFB,BMgr} 
on monolayer graphene
but essentially differs from them in this handling of cutoff-dependent corrections 
by renormalization, which allows one to determine many-body corrections
uniquely from one resonance to another.
Our analysis also reveals that for bilayer graphene 
both intralayer and interlayer coupling 
strengths undergo renormalization.~\cite{fnzero,KCC}
We compare theory with the experimental data for monolayer and bilayer graphene.

In Sec.~II we briefly review the effective theory of graphene and,
in Sec.~III, study cyclotron resonance in monolayer graphene, 
with focus on the many-body corrections and renormalization.
In Sec.~IV we extend our analysis to bilayer graphene.
Section~V is devoted to the summary and discussion.

\section{monolayer graphene}

Graphene has a honeycomb lattice which consists of two triangle
sublattices of carbon atoms.
The electrons in graphene are described 
by a two-component spinor field $(\psi_{A}, \psi_{B})$ 
on two inequivalent lattice sites $(A,B)$.
The electrons acquire a linear spectrum near the two inequivalent Fermi
points ($K$ and $K'$) in the Brillouin zone,
with the $\lq\lq$light velocity" $v_{0} = (\sqrt{3}/2)\, 
a_{\rm L}\gamma_{0}/\hbar \approx 10^{6}$~m/s 
related to the intralayer coupling 
$\gamma_{0} \equiv \gamma_{AB} \approx 2.9$ eV  
(with $a_{\rm L}=  0.246$nm).

Their low-energy features are described by  
an effective Hamiltonian of the form~\cite{Semenoff}
\begin{eqnarray}
H_{0}&=& \int d^{2}{\bf x}\Big[ \psi^{\dag} {\cal H}_{+} \psi 
+ {\psi'}^{\dag} {\cal H}_{-}\psi' \Big], \nonumber\\
{\cal H}_{\pm}&=& v_{0}\,  (\sigma_{1}\Pi_{1} 
+ \sigma_{2}\Pi_{2} \mp \delta m\, \sigma_{3})  - e A_{0}, 
\label{Hzero}
\end{eqnarray}
where $\Pi_{i} = -i\partial_{i} + e A_{i}$ [$i= (1,2)$ or $(x,y)$] 
involve coupling to external electromagnetic potentials
$A_{\mu}= (A_{i}, A_{0})$.
Here $\psi\equiv (\psi_{1},\psi_{2})^{t}=(\psi_{A},\psi_{B})^{t}$ 
stands for the electron field 
at one (or $K$) valley, and $\psi'=(\psi'_{B},\psi'_{A})^{t}$ 
to one at another valley, with $A$ and $B$ 
referring to the associated lattice sites.
$\delta m$ denotes a possible tiny asymmetry in sublattices.

Let us place graphene in a uniform magnetic field $B_{\perp}=B>0$; 
we set  $A_{i} \rightarrow B\, (-y,0)$.
The electron spectrum then forms an infinite tower of 
Landau levels $L_{n}$ of energy 
\begin{equation}
\epsilon_{n} = s_{n}\,  \omega_{c} 
\sqrt{|n|+ (\delta m)^{2}\ell^{2}/2},
\end{equation}
labeled by integers $n=0,\pm 1, \pm2, \dots$, and
$p_{x}$ (or $y_{0} \equiv \ell^{2} p_{x}$ with the magnetic length 
$\ell \equiv 1/\sqrt{eB}$).
Here $s_{n} \equiv {\rm sgn}\{n\} =\pm 1$ specifies 
the sign of the energy $\epsilon_{n}$, and 
\begin{equation}
\omega_{c}\equiv\sqrt{2}\, v_{0}/\ell
\approx 36.3\times v_{0}[10^{6}{\rm m/s}]  \sqrt{B[{\rm T}]}\, {\rm meV}.
\end{equation}
is the basic cyclotron frequency;
$v_{0}[10^{6}{\rm m/s}]$ stands for $v_{0}$ in units of $10^{6}{\rm m/s}$ 
and $B[{\rm T}]$ for a magnetic field in tesla.

Suppose, without loss of generality, that $\delta m>0$. 
Then the $n=0$ level at the $K$ valley has positive energy 
$\epsilon_{0_{+}} = v_{0} \delta m > 0$ while
the $n=0$ level at the $K'$ valley has negative energy 
$\epsilon_{0_{-}} = -v_{0} \delta m$.
In general, the spectra at the two valleys are related as
$\epsilon_{n}|_{K}= -\epsilon_{-n}|_{K'}$.
With the electron spin taken into account 
(and Zeeman splitting ignored for simplicity), 
each Landau level is thus fourfold degenerate, except for 
the doubly-degenerate $n=0_{\pm}$ levels split in valley.
With this feature in mind, we shall set the asymmetry 
$\delta m \rightarrow 0$ in what follows.

The Coulomb interaction is written as
\begin{equation}
H^{\rm Coul} 
= {1\over{2}} \sum_{\bf p}
v_{\bf p}\, :\rho_{\bf -p}\, \rho_{\bf p}:,
\label{Hcoul}
\end{equation}
where $\rho_{\bf p}$ is the Fourier transform of the electron
density $\rho = \psi^{\dag}\psi + {\psi'}^{\dag}\psi'$
(here $\psi^{\dag}\psi$, e.g., is summed over spinor 
and spin indices);
$v_{\bf p}= 2\pi \alpha/(\epsilon_{\rm b} |{\bf p}|)$ is 
the Coulomb potential with the fine-structure constant 
$\alpha = e^{2}/(4 \pi \epsilon_{0}) \approx 1/137$ and 
the substrate dielectric constant $\epsilon_{\rm b}$.

The Landau-level structure is made explicit by passing to
the $|n,y_{0}\rangle$ basis, with the expansion
$\psi ({\bf x}, t) = \sum_{n, y_{0}} \langle {\bf x}| n, y_{0}\rangle\, 
\psi_{n}(y_{0},t)$. (For conciseness, we shall only display the $\psi$
sector from now on.)
The Hamiltonian $H$ is thereby rewritten as
\begin{equation}
H_{0}\! = \!\! \int\! dy_{0} \!\!\!
\sum_{n =-\infty}^{\infty} \!\!\!
\psi^{\dag}_{n} (y_{0},t)\, \epsilon_{n}\, \psi_{n} (y_{0},t),
\label{Hzeronn}
\end{equation}
and the charge density $\rho_{-{\bf p}}(t) =\int d^{2}{\bf x}\,  
e^{i {\bf p\cdot x}}\,\psi^{\dag}\psi$ as~\cite{KSgr}
\begin{eqnarray}
\rho_{-{\bf p}} &=&\sum_{k, n=-\infty}^{\infty} \rho^{k n}_{\bf -p}
=\sum_{k, n=-\infty}^{\infty} g^{k n}_{\bf p}\, 
R^{k n}_{\bf p}, \nonumber\\
R^{kn}_{\bf p}&=& \gamma_{\bf p}\int dy_{0}\,
\psi_{k}^{\dag}(y_{0},t)\, e^{i{\bf p\cdot r}}\,
\psi_{n} (y_{0},t),
\label{chargeoperator}
\end{eqnarray}
where $\gamma_{\bf p} =  e^{- \ell^{2} {\bf p}^{2}/4}$; 
${\bf r} = (r_{x}, r_{y}) = (i\ell^{2}\partial/\partial y_{0}, y_{0})$
stands for the center coordinate with uncertainty 
$[r_{x}, r_{y}] =i\ell^{2}$.
The charge operators $R^{k n}_{\bf p}$ obey the $W_{\infty}$ algebra~\cite{GMP}
\begin{equation}
\ [R^{m\, m'}_{\bf k} , R^{n\, n'}_{\bf p}]
= \delta^{m' n} e^{k^{\dag} p/2 }\,
 R^{m\, n'}_{\bf k+p}
-\delta^{n' m} e^{p^{\dag} k/2 }\,
 R^{n\, m'}_{\bf k+p},
\label{RRalg} 
\end{equation}
where $ k^{\dag} p= {\bf k\! \cdot\! p} 
- i\, {\bf k}\! \times\! {\bf p}$ with ${\bf k}\! \times\! {\bf p}
\equiv k_{x}p_{y} -k_{y}p_{x}$.
This actually consists of two $W_{\infty}$ algebras 
associated with intralevel center-motion~\cite{GMP} 
and interlevel mixing~\cite{KSsma} of electrons.

The coefficient matrix $g^{kn}_{\bf p}$ is given by
\begin{equation}
g^{k n}_{\bf p} = {1\over{2}}\, b_{k}b_{n} \big(
f^{|k|\! -\!1, |n| -\!1}_{\bf p} + s_{k}s_{n}\, 
f^{|k|, |n|}_{\bf p} \big), 
\label{gpsikn}
\end{equation}
where $b_{n}=1$ for $n\not=0$ and $b_{0}=\sqrt{2}$;  
\begin{equation}
f^{k n}_{\bf p} 
= \sqrt{{n!\over{k!}}}\,
\Big({-\ell p\over{\sqrt{2}}}\Big)^{k-n}\, L^{(k-n)}_{n}
\Big(\textstyle{1\over{2}} \ell^{2}{\bf p}^{2}\Big)
\label{fknp}
\end{equation}
for $k \ge n\ge0$, and $f^{n k}_{\bf p} = (f^{k n}_{\bf -p})^{\dag}$;
$p=p_{x}\! -\!ip_{y}$.
Note that $g^{k n}_{\bf p}$ are essentially the same at the two valleys, 
i.e., $g^{k n}_{\bf p}|_{K'}= g^{k n}_{\bf p}|_{K}$ 
and $g^{k 0_{-}}_{\bf p}|_{K'}= g^{k 0_{+}}_{\bf p}|_{K}$ for $|k|\ge 1$ and 
$|n| \ge 1$; one simply needs to specify $n=0_{\pm}$ accordingly.

\section{Cyclotron resonance}

In this section we study cyclotron resonance in monolayer graphene. 
Let us first note that the charge operator $\rho^{kn}_{\bf -p}= g^{k n}_{\bf p}\, 
R^{k n}_{\bf p}$ in Eq.~(\ref{chargeoperator}) annihilates an electron
at the $n$th level $L_{n}$ and creates one at the $k$th level $L_{k}$.
One may thus associate it with the interlevel transition $L_{n}\rightarrow L_{k}$
or regard it as an interpolating operator for the exciton consisting 
of a hole at $L_{n}$ and an electron at $L_{k}$.

To describe such inter-Landau-level excitations one can make use of a nonlinear realization 
of the $W_{\infty}$ algebra, as is familiar from the theory 
of quantum Hall ferromagnet.~\cite{MoonMYGM}
One may start with a given ground state $|G\rangle$ and 
describe an excited collective state $|\tilde{G}\rangle$ over it 
as a local rotation in $(n,y_{0})$ space,
\begin{equation}
|\tilde{G}\rangle = e^{-i{\cal O}} |G\rangle,
\end{equation}
where the operator $e^{-i{\cal O}}$ with
\begin{equation}
{\cal O} = \sum_{\bf p}\gamma_{\bf p}^{-1}\, 
\Phi^{kn}_{\bf p}\, R^{kn}_{\bf p}
\end{equation}
locally rotates $|G\rangle$ by $\lq\lq$angles" $\Phi^{kn}_{\bf p}$, 
which define textures in $(n,y_{0})$ space.
(Remember here that $\rho_{\bf p} \sim R^{kn}_{\bf p}$ are diagonal in spin 
so that $\Phi^{kn}_{\bf p}$ carry the spin index as well, 
though it is suppressed. 
In principle, one has to retain in ${\cal O}$ all possible  
pairs $\Phi^{kn}_{\bf p}$ and  
$(\Phi^{kn}_{\bf p})^{\dag}=\Phi_{\bf -p}^{nk}$ 
contributing to the $L_{n}\rightarrow L_{k}$ transition.)

Repeated use of the charge algebra~(\ref{RRalg})
allows one to express the texture-state energy 
$\langle \tilde{G}| H |\tilde{G}\rangle
=\langle G|e^{i{\cal O}} H e^{-i{\cal O}}|G\rangle$
with $H=H_{0} + H^{\rm Coul}$
as a functional of $\Phi^{kn}_{\bf p}$ or its ${\bf x}$-space 
representative $\Phi^{kn}({\bf x},t)$.
The kinetic term for $\Phi^{kn}$ is supplied 
from the electron kinetic term,
and one can write the effective Lagrangian for $\Phi$ as
\begin{equation}
L_{\Phi}=\langle \tilde{G}|(i \partial_{t} -H) |\tilde{G}\rangle
=\langle G|e^{i{\cal O}}\, 
(i \partial_{t} -H) e^{-i{\cal O}}|G\rangle.
\end{equation}
This representation systematizes 
the single-mode approximation~\cite{GMP}  
(SMA) within a variational framework.~\cite{KSsma}
The present theory thus embodies the nonperturbative 
features of the SMA.

Indeed, for a transition from the filled level $L_{n}$ 
(with density $\bar{\rho}_{n}$) to the empty level $L_{k}$, 
one finds 
\begin{equation}
L_{\Phi} = \bar{\rho}_{n}\sum_{\bf p} \Phi^{nk}_{\bf -p}
(i\partial_{t} -\epsilon^{\rm exc}_{\bf p})\Phi^{kn}_{\bf p} + \cdots,
\label{Lphi}
\end{equation}
with the excitation spectrum given 
by the SMA formula,
\begin{equation}
\epsilon^{\rm exc}_{\bf p} =
\langle G| [\rho^{nk}_{\bf p}, [H,\rho^{kn}_{\bf -p}] ] |G\rangle/
\langle G| \rho^{nk}_{\bf p}\rho^{kn}_{\bf -p} |G \rangle,
\label{Eexcp}
\end{equation}
i.e., as the oscillator strength divided by the static structure factor,
\begin{equation}
\langle G| \rho^{nk}_{\bf p}\rho^{kn}_{\bf -p} |G \rangle /\Omega
= \bar{\rho}_{n}\gamma_{\bf p}^{2}\, |g^{kn}_{\bf p}|^{2},
\end{equation}
where $\Omega = \int d^{2}{\bf x}$.

Let us first consider the $n=0_{\pm}\rightarrow n=1$ transitions
at filling factor $\nu= 2$,
with all $n\le 0_{\pm}$ levels filled and all $n\ge 1$ levels empty.
A laborious direct calculation of Eq.~(\ref{Eexcp}) 
yields $\epsilon^{\rm exc}_{\bf k} = \epsilon_{1} -\epsilon_{0} 
+ \triangle \epsilon_{\bf k}^{10}$ with
\begin{eqnarray}
\triangle \epsilon_{\bf k}^{10}
&=& \bar{\rho}_{0}\, v_{\bf k}\, \gamma_{\bf k}^{2}\, 
|g^{10}_{\bf k}/g^{00}_{\bf k}|^{2} 
+\sum_{\bf p}v_{\bf p} \gamma_{\bf p}^{2}\, I_{{\bf p},{\bf k}},
 \nonumber\\
I_{{\bf p},{\bf k}} &=&
 \sum_{n\le 0} (|g^{0n}_{\bf p}|^{2}- |g^{1n}_{\bf -p}|^{2})
-c_{\bf p,k}\,  g^{11}_{\bf p}g^{00}_{\bf -p},\
\label{deltaEk}
\end{eqnarray}
where $\gamma_{\bf k}^{2} = e^{-  \ell^{2}{\bf k}^2/2}$
and $c_{\bf p,k}=\cos (\ell^{2}{\bf p\! \times\! k})$.

When the $n=0_{\pm}$ level is partially filled, 
$\triangle \epsilon^{10}_{\bf k}$ involves the following contribution,
\begin{eqnarray}
\delta\epsilon^{10}_{\bf k}&=&\sum_{\bf p}v_{\bf p}\,\gamma_{\bf p}^{2}
\Big[  |g^{01}_{\bf p}/g^{00}_{\bf p}|^{2}\,  
\bar{s}({\bf p+k})/\gamma_{\bf p+k}^{2}
\nonumber\\
&&+ ( c_{\bf p,k}\, g^{11}_{\bf p}/g^{00}_{\bf p} - 1)\, 
 \bar{s}({\bf p})/\gamma_{\bf p}^{2} \Big], 
\label{partfilled}
\end{eqnarray}
where $\bar{s}({\bf p})$ stands for the projected static structure factor
defined as 
$\langle G| \rho^{00}_{\bf - p}\rho^{00}_{\bf p} |G\rangle /\Omega
= \bar{\rho}_{0}\{\bar{\rho}_{0}\delta_{{\bf p},0} +\bar{s}({\bf p})\}$.  
The determination of $\bar{s}({\bf p})$ is a highly nontrivial task
which requires an exact diagonalization study of model systems,~\cite{AA} 
and is not attempted here.  We instead focus on the case of integer filling,
for which  $\bar{s}({\bf p})$ is taken to vanish.

Equation~(\ref{deltaEk}), 
together with Eq.~(\ref{partfilled}), 
essentially agrees with the result of MacDonald and Zhang~\cite{MZ}
for the $L_{0} \rightarrow L_{1}$ transition in the standard QH system. 
The key difference is that quantum fluctuations in $I_{{\bf p},{\bf k}}$
now involve a sum $\sum_{n\le -1}(\cdots)$ 
over infinitely many Landau levels in the valence band (or the Dirac sea).
Actually, one can verify that 
the SMA expressions~(\ref{deltaEk}) and~(\ref{partfilled}) equally 
apply to a general interlevel transition
$L_{a} \rightarrow L_{b}$
if one sets the superscripts $0\rightarrow a$ and $1\rightarrow b$, 
in an obvious fashion, and takes the sum $\sum_{n}$ over filled levels.

To be precise, the $0\rightarrow 1$ transition of our interest consists of 
four channels, $(0_{+}\rightarrow 1)|_{K}$  and $(0_{-}\rightarrow 1)|_{K'}$
each with spin $s_{z}=\pm 1/2$.
One therefore has to consider mixing of four $\Phi^{{10}}$
to determine $\epsilon^{\rm exc}_{\bf k}$.
Actually only the first term $\propto v_{\bf k}\,|g^{10}_{\bf k}|^{2}$ 
in Eq.~(\ref{deltaEk}),
which comes from the direct Coulomb exchange, is responsible 
for such mixing~\cite{BMgr,IWFB}
because the rest of terms are diagonal in spin and valley.
Such direct terms $\propto v_{\bf k}\,|g^{ba}_{\bf -k}|^{2}$
(with $b\not=a$) in general vanish for ${\bf k}\rightarrow 0$,
and mixing thus takes place only for ${\bf k}\not=0$.
In what follows we focus on the ${\bf k}=0$ excitation energies 
$\epsilon^{\rm exc}_{{\bf k}=0}$, with no mixing taken into account.
In addition, we make no distinction between the $0_{\pm}$ levels 
because $g^{kn}_{\bf p}$ are essentially the same at the two valleys, 
as noted in Sec.~II.

The cyclotron-resonance energy 
for a general $L_{a}\rightarrow L_{b}$ transition 
with the Landau levels filled up to $n=n_{\rm f}$
is written as $\epsilon^{\rm exc}_{{\bf k}=0} = \epsilon_{b} -\epsilon_{a} 
+ \triangle \epsilon_{{\bf k}=0}^{b\leftarrow a}$ with
\begin{equation}
\triangle \epsilon^{b\leftarrow a}_{{\bf k}= 0} = \sum_{\bf p}v_{\bf p}\,
\gamma_{\bf p}^{2}\, \Big[
\sum_{n\le n_{\rm f}} (|g^{an}_{\bf -p}|^{2}- |g^{b n}_{\bf p}|^{2})
- g^{bb}_{\bf p}\, g^{aa}_{\bf -p}\Big].
\nonumber\\
\label{energyshift}
\end{equation}
This is the basic formula we use in what follows.
Note that $n_{\rm f}=1,0_{+},-1,-2, -3, ...$ 
correspond to the filling factors 
$\nu = 4\,n_{\rm f}+ 2= 6, 2,-2,-6,-10, ...$,
respectively.
The $\sum_{n\le n_{\rm f}} (|g^{an}_{\bf p}|^{2}- |g^{bn}_{\bf -p}|^{2})$ term
refers to the change in quantum fluctuations, 
via the $a\rightarrow b$ transition, of the filled states.
Its structure is easy to interpret physically:
As an electron is excited from $L_{a}$ to $L_{b}$, 
$|n=a,y_{0}\rangle \rightarrow |n=b,y'_{0}\rangle$, 
virtual transitions from any filled levels 
to the $|b,y'_{0}\rangle$ state are forbidden 
while those to the newly unoccupied $|a,y_{0}\rangle$ state
are allowed to start.

For standard QH systems
this correction $\triangle \epsilon^{b\leftarrow a}_{{\bf k}=0}$
vanishes for each transition to the adjacent level,
$L_{n}\rightarrow L_{n+1}$, according to Kohn's theorem.~\cite{Kohn}
Indeed, one can verify, for the $0\rightarrow 1$ transition,
the relation
\begin{equation}
|g^{00}_{\bf -p}|^{2}- |g^{10}_{\bf p}|^{2}
- g^{11}_{\bf p}\, g^{00}_{\bf -p}=0
\label{kohnth}
\end{equation}
(with $g^{00}_{\bf p}\rightarrow 1$ and 
$g^{10}_{\bf p}\rightarrow -\ell\, p/\sqrt{2}$)
and analogous ones for other $L_{n}\rightarrow L_{n+1}$ as well.

Interestingly,  it happens that 
Eq.~(\ref{kohnth}) also holds for the $0_{\pm}\rightarrow 1$ transition in graphene, 
with $g^{00}_{\bf p}=1$, $g^{10}_{\bf p}=-\ell\, p/2$ and
$g^{11}_{\bf p}=1 - \ell^{2}{\bf p}^{2}/4$.
Any nonzero shift $\triangle \epsilon^{1 \leftarrow 0}_{{\bf k}=0}$ 
for the $0\rightarrow 1$ cyclotron resonance therefore
comes from the quantum fluctuations of the Dirac sea, and  
actually diverges logarithmically  
with the number  $N_{\rm L}$ of filled Landau levels in the sea,
\begin{eqnarray}
\triangle \epsilon^{1\leftarrow 0}_{\bf 0}\! &=&\! \sum_{\bf p}v_{\bf p}\,
\gamma_{\bf p}^{2}\!\!\!
\sum_{-N_{L}\le n\le -1} \!\!\!\! (|g^{0n}_{\bf -p}|^{2}- |g^{1n}_{\bf p}|^{2})
=V_{c}\, {\cal C}_{N},  \nonumber\\
{\cal C}_{N}&\approx& (\sqrt{2}/8)\, (\, \log N_{L} - 1.017),
\label{logNL}
\end{eqnarray}
where 
\begin{equation}
V_{c} \equiv \alpha/(\epsilon_{b}\ell) \approx (56.1/\epsilon_{b})\, 
\sqrt{B[{\rm T}]}\, {\rm meV}.
\end{equation}
${\cal C}_{N}$ agrees~\cite{fnone} with the result of earlier works~\cite{IWFB,BMgr} 
obtained by a different method.

This divergence in ${\cal C}_{N}$ derives from short-wavelength vacuum polarization
and is present even for $B=0$.
To see this one may evaluate the Coulomb exchange correction in free space
(with $\alpha\rightarrow \alpha/\epsilon_{b}$), 
using the instantaneous photon and fermion propagators
$v_{\bf k} =2\pi\, \alpha/(\epsilon_{b} |{\bf k}|)$ and 
$iS({\bf p}) = \sigma_{i}p_{i} /(2\, |{\bf p}|)$,
\begin{equation}
\sum_{\bf k}v_{\bf k}\, iS({\bf p+k})
= \sigma\! \cdot\! {\bf p}\,{\alpha\over{8 \epsilon_{b}}}\, 
\log (C\, \Lambda^{2}/{\bf p}^{2}), 
\label{vS}
\end{equation}
with momentum cutoff $|{\bf k}|\le \Lambda$ and some constant $C$.

This divergent correction causes infinite renormalization~\cite{velrenorm} 
of velocity $v_{0}$
in the electron kinetic term $\propto v_{0}\sigma ^{i}\partial_{i}$;
Eq.~(\ref{vS}) agrees with an earlier result of Ref.~\onlinecite{velrenorm}.
It vanishes at ${\bf p}=0$ 
but, for $B\not =0$, turns into a nonvanishing energy gap, 
with ${\bf p}^{2}\rightarrow 2eB= 2/\ell^{2}$.
Actually, this diverging piece precisely agrees with that 
in Eq.~(\ref{logNL}), if one simply chooses 
the $\lq\lq$Fermi momentum" $\Lambda$ so that the Dirac sea accommodates 
the same number of electrons as in the $B\not=0$ case,
$N_{\rm sea} = \Lambda^{2}/4\pi \approx N_{L}/2\pi \ell^{2}$, i.e., 
$\Lambda^{2}\approx 2N_{L}/\ell^{2}$.

Since such an infinite correction is already present for $B=0$,
it does not make sense to discuss the magnitude 
of the cutoff-dependent number ${\cal C}_{N}$ in Eq.~(\ref{logNL}).
The legitimate procedure is to renormalize $v_{0}$
by rescaling 
\begin{equation}
v_{0}=Z_{v}\, v_{0}^{\rm ren}
\end{equation}
and put reference 
to the cutoff into $Z_{v}$, with $v_{0}^{\rm ren}$ regarded 
as an observable quantity.

The renormalized velocity $v_{0}^{\rm ren}$ is defined 
by referring to a specific resonance.
Let us take the $0\rightarrow 1$ resonance and choose to absorb 
the entire $O(V_{c})$ correction at some reference scale
(e.g., at magnetic field $B_{0}$) into $Z_{v}$, i.e., 
we write
\begin{equation}
\epsilon^{1\leftarrow 0}_{{\bf k}= 0} \stackrel{\nu=2}{=}
\epsilon_{1} + \triangle \epsilon_{{\bf k}=0}^{1\leftarrow 0}
= \sqrt{2}\, v_{0}^{\rm ren}|_{B}/\ell
\equiv  \omega_{c}^{\rm ren}|_{B}
\end{equation}
by setting
\begin{equation}
Z_{v} = 1- {\alpha\over{\sqrt{2}v_{0}\epsilon_{b}}}\, {\cal C}_{N}
=1- {\alpha\over{8v_{0}\epsilon_{b}}}\, 
\log\, {\Lambda^{2}\over{\kappa^{2}}},
\end{equation}
where $\kappa^{2} = ({\rm const.})\, eB_{0}$.
The renormalized velocity then depends on $B$, or runs with $B$, 
\begin{equation}
v_{0}^{\rm ren}|_{B} = 
v_{0}^{\rm ren}|_{B_{0}} - {\alpha\over{8\epsilon_{b}}}\, \log (B/B_{0}), 
\end{equation}
decreasing slightly for $B>B_{0}$; actually the correction is 
rather small (about 3\% for $B/B_{0}\sim 2$ and $\epsilon_{b} \sim 5$, 
with $v_{0}^{\rm ren} \sim c/300$).
With such $B$ dependence in mind, we denote 
$v_{0}^{\rm ren}|_{B}$ as $v_{0}^{\rm ren}$ 
and $\omega_{c}^{\rm ren}|_{B}$
as $\omega_{c}^{\rm ren}$ from now on.


\begin{figure}[tbp]
\includegraphics[scale=0.225]{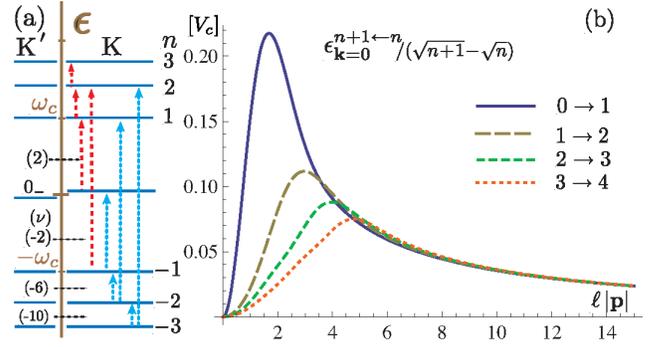}     
\caption{
(a) Cyclotron resonance; circularly-polarized light can distinguish 
between two classes of transitions indicated by different types of arrows.
(b) Momentum profiles of the many-body corrections 
$\triangle \epsilon^{n+1\leftarrow n}_{{\bf k}=0}/(\sqrt{n+1} -\sqrt{n})$
in units of $V_{c}$ for $n=$ 0,1,2 and 3.
}
\end{figure}


The divergences in $\epsilon^{\rm exc}_{\bf k} =\epsilon_{k} -\epsilon_{n} 
+ \triangle \epsilon_{\bf k}$ for all other resonances,
as illustrated in Fig.~1~(a), are taken care of 
by this velocity renormalization.
The finite corrections $\propto V_{c}$ 
after renormalization then make sense as genuine observable corrections.
In particular, for several intraband channels 
 $L_{n}\rightarrow L_{n+1}$ at filling factor 
$\nu= 4\, n_{\rm f} + 2$, direct calculations yield
\begin{eqnarray}
\epsilon^{2\leftarrow 1}_{{\bf k}= 0}&\stackrel{\nu=6}{=}& (\sqrt{2}-1)\, 
\big\{ \omega_{c}^{\rm ren} - 0.264\, V_{c} \big\} ,\nonumber\\
\epsilon^{3\leftarrow 2}_{{\bf k}= 0}&\stackrel{\nu=10}{=}& (\sqrt{3}-\sqrt{2})\, 
\big\{ \omega_{c}^{\rm ren} - 0.358\, V_{c} \big\} ,\nonumber\\
\epsilon^{4\leftarrow 3}_{{\bf k}= 0}&\stackrel{\nu=14}{=}& (\sqrt{4}- \sqrt{3})\,
\big\{ \omega_{c}^{\rm ren} - 0.419\, V_{c}\big\} ,\nonumber\\
\epsilon^{5\leftarrow 4}_{{\bf k}= 0}&\stackrel{\nu=18}{=}& (\sqrt{5}- \sqrt{4})\, 
\big\{ \omega_{c}^{\rm ren} - 0.464\, V_{c} \big\} ,
\label{CRintraband}
\end{eqnarray}
where $V_{c}= \alpha/(\epsilon_{b}\ell)$.
The Coulomb corrections, shown numerically here, 
are analytically calculable. 
The excitation spectra $\epsilon_{\bf k}$
in the hole band are essentially the same,
\begin{equation}
\epsilon^{-n\leftarrow -(n+1)}_{\bf k}|_{n_{\rm f}=-(n+1)}^{\nu=-(4n +2)}
=\epsilon^{n+1\leftarrow n}_{\bf k}|_{n_{\rm f}=n}^{\nu=4n +2},
\end{equation}
reflecting the particle-hole symmetry.

Figure 1~(b) shows some of 
the momentum profiles $\gamma_{\bf p}^{2}\, [\cdots]$ 
in $\triangle \epsilon^{n+1\leftarrow n}_{{\bf k}=0}$ of Eq.~(\ref{energyshift}),
which, when integrated over $\ell |{\bf p}|$, give
$\triangle \epsilon^{n+1\leftarrow n}_{{\bf k}=0}/(\sqrt{n+1} -\sqrt{n})$
in units of $V_{c}$.
It is clearly seen that the slowly decreasing high-momentum tails
$\sim (\sqrt{2}/4)/(\ell\, |{\bf p}|)$ 
are responsible for the ultraviolet (UV) divergence and that 
the finite observable corrections $\propto V_{c}$ are uniquely determined 
from the profiles in the low-momentum region $\ell\, |{\bf p}| \lesssim 15$.

A look into the structure of the total current operator tells us that
the optically-induced cyclotron resonance (for ${\bf k}=0$) in graphene 
is governed by the selection rule $\triangle |n|=\pm 1$, 
in contrast to the $\lq\lq$nonrelativistic" rule
$\triangle n=\pm 1$.
In particular, there are two classes of transitions, 
(i)~$-n \rightarrow \pm (n-1)$ and (ii)~$\pm (n-1)\rightarrow n$ (with $n\ge 1$),
which are distinguished~\cite{AFAC} 
by use of circularly-polarized light ($\propto A_{x} \pm i A_{y}$); see Appendix B.

As a result, graphene supports interband cyclotron resonances.
The lowest channels are open at $\nu=-2$, with 
\begin{eqnarray}
\epsilon^{2 \leftarrow -1}_{{\bf k}= 0}&\stackrel{\nu=-2}{=}&(\sqrt{2}+1)\, 
\big\{ \omega_{c}^{\rm ren} + 0.122\, V_{c} \big\} ,
\nonumber\\
\epsilon^{1\leftarrow-2}_{{\bf k}= 0}&\stackrel{\nu=-2}{=}&(\sqrt{2}+1)\, 
\big\{ \omega_{c}^{\rm ren} + 0.155\, V_{c} \big\} .
\end{eqnarray}
Some other interband channels yield
\begin{eqnarray}
\epsilon^{1\leftarrow-2}_{{\bf k}= 0}&\stackrel{\nu=-6}{=}&(\sqrt{2}+1)\, 
\big\{ \omega_{c}^{\rm ren} + 0.084\, V_{c} \big\},
\nonumber\\
\epsilon^{3\leftarrow-2}_{{\bf k}= 0}&\stackrel{\nu=-6}{=}&(\sqrt{2}+\sqrt{3})\, 
\big\{ \omega_{c}^{\rm ren} + 0.058\, V_{c} \big\},
\nonumber\\
\epsilon^{2\leftarrow-3}_{{\bf k}= 0}&\stackrel{\nu=-6}{=}&(\sqrt{2}+\sqrt{3})\, 
\big\{ \omega_{c}^{\rm ren} + 0.114\, V_{c} \big\},
\nonumber\\
\epsilon^{2\leftarrow-3}_{{\bf k}= 0}&\stackrel{\nu=-10}{=}&(\sqrt{2}+\sqrt{3})\, 
\big\{ \omega_{c}^{\rm ren} + 0.044\, V_{c} \big\}.
\label{CRinterband}
\end{eqnarray}

It is now clear that cyclotron resonance is best analyzed by plotting
the rescaled energies  $\epsilon^{b\leftarrow a}_{{\bf k}= 0}
/|s_{b}\sqrt{|b|} - s_{a}\sqrt{|a|}|$ as a function of $\sqrt{B}$ or $B$.
The Coulombic many-body effect will be
seen as a variation in the characteristic velocity $v_{0}^{\rm ren}[1 + O(V_{c})]$
from one resonance to another, and 
a deviation of $\omega_{c}^{\rm ren}$ from the $\sqrt{B}$ behavior
would indicate the running of $v_{0}^{\rm ren}$ with $B$.


\begin{figure}[tbp]
\includegraphics[scale=0.36]{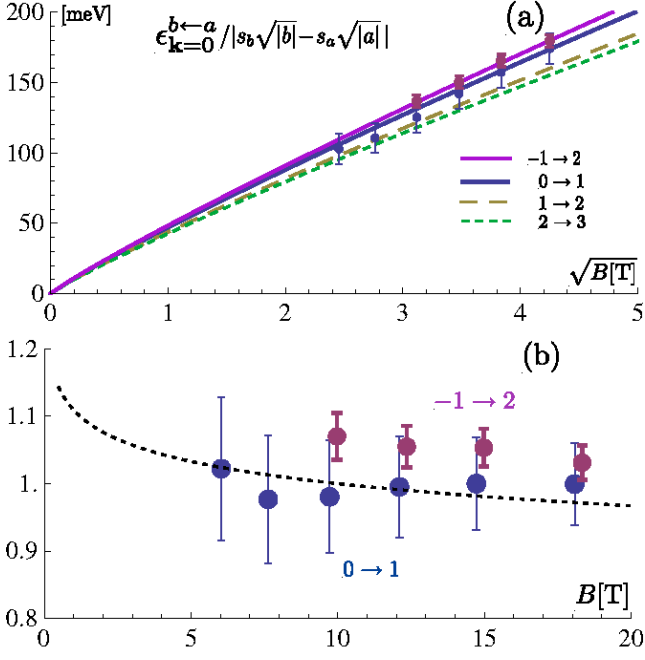} 
\caption{(a) Rescaled cyclotron-resonance energies as a function of $\sqrt{B}$,
with $v_{0}^{\rm ren}\approx 1. 13 \times 10^{6}$ m/s
(at $B=10$\,T) and $V_{c} \approx 12\, 
\sqrt{B[{\rm T}]}\, {\rm meV}$  ($\epsilon_{b}\sim 5$).
The experimental data on $\epsilon^{1\leftarrow 0}_{{\bf k}= 0}$
and $\epsilon^{2\leftarrow -1}_{{\bf k}= 0}/(\sqrt{2}+1)$ are quoted 
from Ref.~\onlinecite{JHT}, 
with error bars inferred from the symbol size in the original data.
(b) The same data plotted in units of 
$\omega_{c}=\sqrt{2}\,v_{0}/\ell$
(with $v_{0}\rightarrow v_{0}^{\rm ren}|_{B=10{\rm T}}$) 
as a function of $B$.
The dotted curve represents a possible 
profile of the running of $v_{0}^{\rm ren}|_{B}$, 
normalized to 1 at $B=10$T, with 
$v_{0}^{\rm ren}|_{B=10{\rm T}}\approx 1. 13 \times 10^{6}$ m/s
and $\epsilon_{b}\approx 5$.
}
\end{figure}


Figure 2 (a) shows such plots for some intra- and inter-band channels,
using $v_{0}^{\rm ren}|_{B=10{\rm T}}\approx 1. 13 \times 10^{6}$ m/s
which fits the $0\rightarrow 1$ resonance data, and,
 as a typical value, 
$V_{c}= \alpha/(\epsilon_{b}\ell) \approx 12\, 
\sqrt{B[{\rm T}]}\, {\rm meV}$  ($\epsilon_{b}\sim 5$).

Actually, experiment~\cite{JHT} has already observed a small deviation 
of the $1: (1+\sqrt{2})$ ratio of 
$\epsilon^{1\leftarrow 0}_{{\bf k}= 0}$
to $\epsilon^{2\leftarrow -1}_{{\bf k}= 0}$
well outside of the experimental errors 
under high magnetic fields 
$B=(6 \sim 18)$ T; the data are apparently electron-hole symmetric,
$\epsilon^{1\leftarrow 0}_{{\bf k}= 0} 
\approx \epsilon^{0\leftarrow -1}_{{\bf k}= 0}$.  
Figure~2~(a) includes such data reproduced from Ref.~\onlinecite{JHT}.
A small increase of $v_{0}^{\rm ren}$
in $\epsilon^{2\leftarrow -1}_{{\bf k}= 0}/(\sqrt{2} +1)$, 
relative to $\epsilon^{1\leftarrow 0}_{{\bf k}= 0}$, 
is roughly consistent with Eq.~(\ref{CRinterband}) 
which suggests a $ 0.122\ V_{c}/\omega_{c}^{\rm ren}\sim 4\%$ increase 
in $v_{0}^{\rm ren}$
(since $V_{c}/\omega_{c}^{\rm ren} \sim 0.3$).

This feature is clearer from Fig.~2~(b), which 
plots the $\epsilon^{1\leftarrow 0}_{{\bf k}= 0}$ 
and $\epsilon^{2\leftarrow -1}_{{\bf k}= 0}/(\sqrt{2} +1)$ 
data as a function of $B$ in units of $\omega_{c}=\sqrt{2}\,v_{0}/\ell \propto\sqrt{B}$
(with $v_{0}\rightarrow v_{0}^{\rm ren}|_{B=10{\rm T}}$). 
The deviation of the $(-1\rightarrow 2)$ resonance data is more pronounced.
In the figure a dotted curve represents a possible 
profile of the running of $v_{0}^{\rm ren}$ with $B$, 
and, especially, the $(-1\rightarrow 2)$ data 
(with smaller error bars) suggests such running.

It is too early to draw any definite conclusion 
from the present data alone, 
but the data is certainly consistent 
(in sign and magnitude) with the present estimate 
of the many-body effect. 
In this connection, let us note that 
an earlier experiment on thin epitaxial graphite~\cite{SMP} also observed 
the $(0\rightarrow 1)$ and $(-1\rightarrow 2)$ resonances,
with apparently no deviation from the $1: (1+\sqrt{2})$ ratio.
This measurement was done under relatively weak magnetic fields
$B=(0.4\sim4)$ T, 
and it could be that a small deviation, under larger error bars, 
simply escaped detection, apart from the potential difference 
between thin graphite and graphene.

More precise measurements of cyclotron resonance, especially 
in the high $B$ domain where the Coulomb interaction becomes sizable,
would be required to pin down the many-body effect in graphene. 
In this respect, the comparison between interband and intraband resonances 
from the {\it same} initial state, e.g.,  $-n \rightarrow \pm (n-1)$ 
at $\nu=2-4n$ with $n=2,3,...$,
would provide a clearer signal for the many-body effect, with the influence 
of other possible sources reduced to a minimum. 
From Eqs.~(\ref{CRintraband}) -~(\ref{CRinterband}) one can read off the variations 
in $v_{0}^{\rm ren}$, 
\begin{eqnarray}
\triangle R(-2 \rightarrow \pm 1) &\stackrel{\nu=-6}{\approx}&
0.34\, V_{c}/\omega_{c}^{\rm ren}  \sim 10\%,
\nonumber\\
\triangle R(-3 \rightarrow \pm 2)&\stackrel{\nu=-10}{\approx}&
0.40\, V_{c}/\omega_{c}^{\rm ren}  \sim 12\%,
\label{ratioML}
\end{eqnarray}
which imply that a comparison of the $(-2 \rightarrow  \pm 1)$ resonances 
and that of the $(-3 \rightarrow  \pm 2)$ resonances 
would find variations in $v_{0}$, about 3 times larger than the $\sim 4\, \%$ variation 
for $\epsilon^{2 \leftarrow -1}_{{\bf k}= 0}/(\sqrt{2}+1)$
{\it vs} $\epsilon^{0 \leftarrow -1}_{{\bf k}= 0}$ at $\nu=-2$.

\section{cyclotron resonances in bilayer graphene}

In this section we consider cyclotron resonance in bilayer graphene.
In bilayer graphene the electrons are described 
by four-component spinor fields on the four inequivalent sites 
$(A,B)$ and $(A',B')$ in the bottom and top layers, arranged 
in Bernal $A'B$ stacking.
Interlayer coupling~\cite{ZLBF} 
$\gamma_{1} \equiv \gamma_{A'B}\sim (0.3-0.4)$ eV
modifies the intralayer linear spectra $\pm v_{0}\, |{\bf p}|$
to yield, in the low-energy branches $|\epsilon| <\gamma_{1}$,
quasiparticles with a parabolic dispersion.~\cite{MF}
They, in a magnetic field, lead to a particle-hole symmetric tower of  
Landau levels $\{L_{n}\}$ 
($n=0_{\pm}, \pm1, ...$) with spectrum,
\begin{equation}
\epsilon_{n} = s_{n}\, \omega_{c}\, 
\eta_{n}( (\gamma_{1}/\omega_{c})^{2}), 
\label{EnBL}
\end{equation}
where $\eta_{n}(x)
=\big\{ (a_{n} + x 
-\sqrt{x^{2} + 2\, a_{n}\, x + 1})/2 \big\}^{1/2}$ with
$a_{n}= 2|n| -1$;  see Appendix A.
The sequence of low-lying levels is made clearer in the form
\begin{equation}
\epsilon_{n} = s_{n}\, \omega^{\rm bi}\,  
\sqrt{|n|(|n|-1)}\, /\xi_{n}(w), 
\label{EnXi}
\end{equation}
with the characteristic cyclotron energy 
\begin{equation}
\omega^{\rm bi} \equiv \omega_{c}^{2}/\gamma_{1}
= 2v_{0}^{2}/(\gamma_{1}\ell^{2})  \sim 5 \, B[{\rm T}]\, {\rm meV},
\end{equation}
where 
$\xi_{n}(w) = \big\{ (1\! +\! a_{n}\, w 
+\sqrt{1 + 2\, a_{n}\, w + w^{2}})/2 \big\}^{1/2}$
and
$w \equiv (\omega_{c}/\gamma_{1})^{2}
=\omega_{c}^{\rm bi}/\gamma_{1} \sim 0.01\, B[{\rm T}] <1$;
$\xi_{n}(0) =1$.
The high-energy branches  $|\epsilon| >\gamma_{1}$ of the spectra 
give rise to another tower of Landau levels,
with spectrum
\begin{equation}
\epsilon^{+}_{n} = s_{n}\, \gamma_{1}\, \xi_{n}(w), 
\label{highenegyBr}
\end{equation}
where $n = \pm1,\pm 2, \cdots$.

Note that $\xi_{n}(w) = 1 + O(eB/\gamma_{1}) $. 
As a result, $\epsilon_{n}$ rises linearly with $B$ 
at low energies $|\epsilon_{n}|<\gamma_{1}$
and turns into a $\sqrt{B}$ rise for $|\epsilon_{n}|\gg\gamma_{1}$. 
Both $\epsilon_{n}$ and $\epsilon^{+}_{n}$ 
approach $\pm \sqrt{|n|}\, \omega_{c}$ for $|n|\rightarrow \infty$,
since the bilayer turns into two isolated layers at short wavelengths.

In the bilayer there arise four zero-energy levels $\epsilon_{n}=0$ 
with $n=(0_{\pm}, \pm 1)$  per spin.
At one valley (say, $K$) they are electron levels with $n=(0_{+},1)$ 
and, at another valley, they are hole levels with $n=(0_{-},-1)$; 
this feature is made explicit with a weak layer asymmetry, 
such as an interlayer voltage
which opens up a (tunable) band gap.~\cite{OBSHR,Mc,CNMPL,OHL} 
With a nonzero band gap, the zero-energy levels evolve into
two quartets of nearly-degenerate levels (separated by the gap),
i.e.,$\lq\lq$pseudo"-zero-mode levels, 
which are expected to support pseudospin waves~\cite{BCNM,KSpzm}
as characteristic collective excitations.

For simplicity, we here turn off such a layer asymmetry 
as well as Zeeman splitting and the effect of trigonal warping 
(coming from $\gamma_{3} \equiv \gamma_{AB'} < \gamma_{1}$).
In view of the small layer separation, we do not distinguish 
between the intralayer and interlayer Coulomb interactions.
Each Landau level $L_{n}$ is thus treated as fourfold degenerate,
except for the zero-mode levels $(L_{0_{+}}, L_{1})$ or 
$(L_{0_{-}}, L_{-1})$ which are fourfold degenerate at each valley.

The effective Hamiltonian for the electrons in bilayer graphene takes 
a $4\times 4$ matrix form which,
for studying the properties of the low-lying levels,
may be reduced to an approximate $2\times 2$ form.~\cite{MF}
Actually, the bending of the spectrum $\epsilon_{n}$  with $B$ is appreciable 
in the high-$B$ domain, $B=(10 \sim 20)$ T, 
where cyclotron resonance in bilayer graphene has been studied experimentally. 
Accordingly we employ the full 4-component spinor description of the bilayer system; 
see Appendix A for details.

The charge density $\rho_{\bf -p}$ (for each spin and valley) takes the same form as 
Eq.~(\ref{chargeoperator}), with $g^{kn}_{\bf p}$ replaced by
\begin{eqnarray}
g^{kn}_{\bf p}&=& D_{k}D_{n}\big[ f^{|k|, |n|}_{\bf p} 
+ \beta_{k}^{(2)}\, \beta_{n}^{(2)}\, f^{|k| -2, |n| -2}_{\bf p}\nonumber\\
&& + (\beta_{k}^{(3)}\, \beta_{n}^{(3)}+\beta_{k}^{(4)}\, \beta_{n}^{(4)})
f^{|k| -1, |n| -1}_{\bf p}\big];
\end{eqnarray}
see Appendix A for the coefficients $\{\beta_{n}^{(i)}\}$ and $D_{n}$.
The sets $(\epsilon_{n}, g^{kn}_{\bf p})$ at the $K$ and $K'$ valleys 
are related as 
\begin{equation}
\epsilon_{n}|_{K'} = - \epsilon_{-n}|_{K}, \ 
g^{kn}_{\bf p}|_{K'} = g^{-k,-n}_{\bf p}|_{K}.
\end{equation}
Actually, for zero band gap,  $g^{kn}_{\bf p}$ are essentially 
the same at the two valleys since one further finds that
\begin{equation}
g^{kn}_{\bf p}|_{K'} = g^{kn}_{\bf p}|_{K}, \ 
g^{k,-1}_{\bf p}|_{K'} = g^{k,1}_{\bf p}|_{K},
g^{k,0_{-}}_{\bf p}|_{K'} = g^{k,0_{+}}_{\bf p}|_{K},
\label{symmetrygkn}
\end{equation}
for $|k|\ge 2$ and $|n| \ge 2$.

One can now use the SMA formula~(\ref{energyshift}) 
to calculate the interlevel excitation energies 
$\epsilon^{\rm exc}_{\bf k} =\epsilon_{b} -\epsilon_{a} 
+ \triangle \epsilon^{b\leftarrow a}_{\bf k}$. 
The result applies to both valleys if one specifies 
the zero-mode levels accordingly.
It is important to remember
that for bilayer graphene the sum $\sum_{n}$ over filled levels involves 
two branches $\epsilon_{-n}$ and $\epsilon^{+}_{-n}$
in the valence band.

\begin{figure}[tbp]
\includegraphics[scale=0.19]{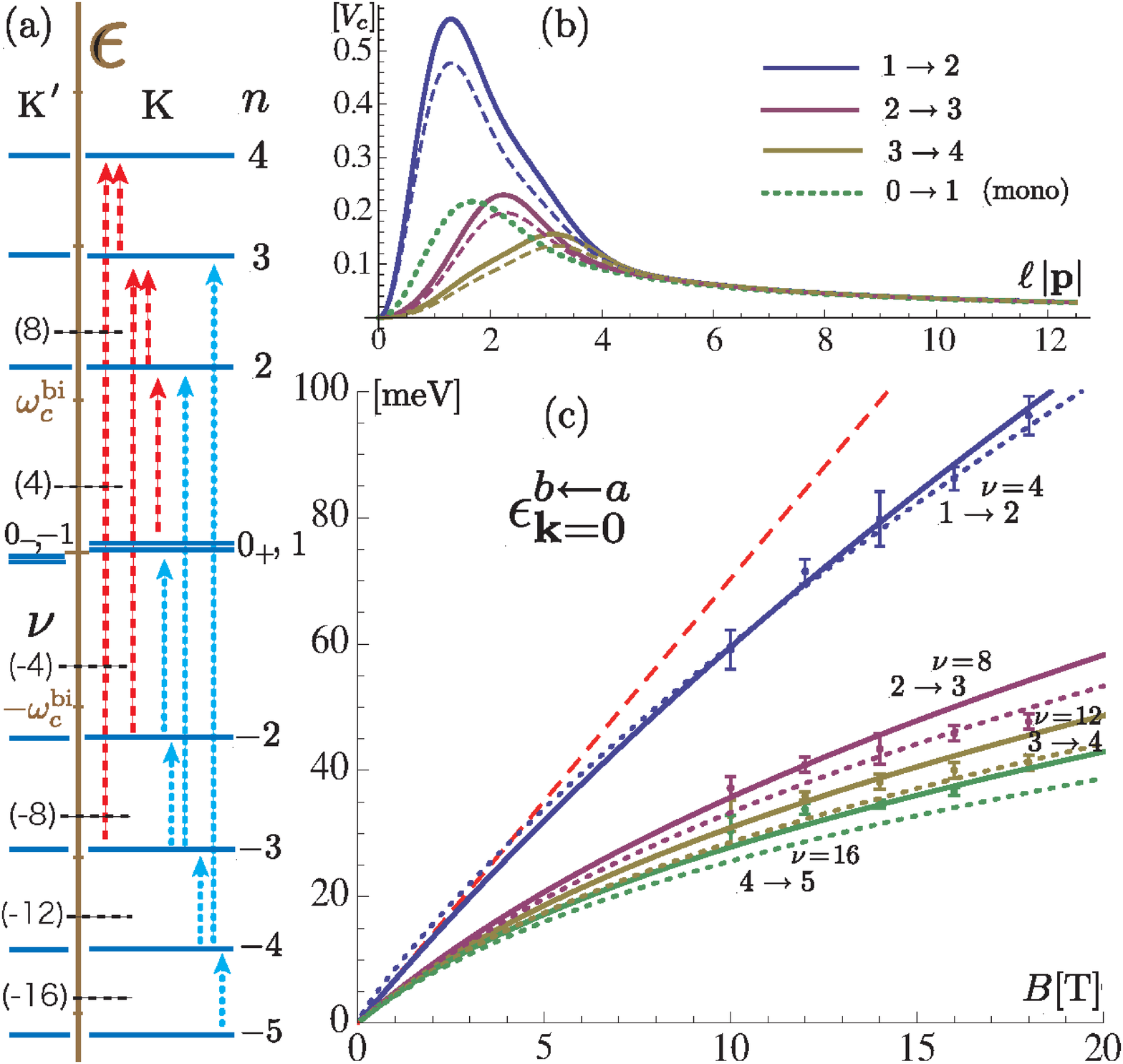} 
\caption{
(a) Cyclotron resonance in bilayer graphene.  
(b)~Momentum profiles of the bilayer many-body corrections 
$\triangle \epsilon^{n+1\leftarrow n}_{{\bf k}=0}/ (\eta_{n+1} -\eta_{n})$
for $n=1,2,$ and 3, with $\hat{v}_{0}=1.15$ and $\hat{\gamma}_{1}=3.5$;
 real curves at $B=$10T and dashed curves at $B=$16T.
A dotted curve refers to the profile of the monolayer $(0\rightarrow 1)$ 
resonance.
(c) Resonance energies as a function of $B$;
real curves, with $\hat{v}_{0}=1.15$, $\hat{\gamma}_{1}=3.5$ and $V_{c}=0$;
dotted curves, with $\hat{v}_{0}=1.15$, $\hat{\gamma}_{1}=3.8$ 
and $V_{c}\approx 5.6\, \sqrt{B[{\rm T}]}$ meV
(or $\epsilon_{b}\approx 10$).
The experimental data with error bars are reproduced from Ref.~\onlinecite{HJTS}.
Note that the low-lying $n=2$ spectrum $\epsilon_{2}$ 
(= the $\nu=4$ curve) significantly deviates from 
the approximate spectrum 
$\epsilon_{2}\approx \sqrt{2}\, \omega_{c}^{\rm bi}$ with $\xi_{2}(w)\rightarrow 1$
(dashed line $\propto B$) for $B>10$T.}
\end{figure}


Cyclotron resonance in bilayer graphene again obeys 
the selection rule~\cite{AFAC} $\triangle |n| = \pm 1$; see Appendix~B
and Fig.~3~(a).
The Coulombic corrections
$\triangle \epsilon^{b\leftarrow a}_{{\bf k}=0}$
are diagonal in spin and valley (while mixing arises for ${\bf k}\not=0$).
The vacuum polarization effect 
again makes $\triangle \epsilon^{b\leftarrow a}_{{\bf k}=0}$
cutoff-dependent.

For renormalization let us first look into the $B=0$ case.
One can construct the electron propagator 
and, as in the monolayer case, calculate the Coulombic quantum corrections.
It turns out that not only $v_{0}$ but also $\gamma_{1}$ undergo infinite 
renormalization and, rather unexpectedly,
the divergent terms are the same for both of them to $O(V_{c})$ at least; 
they also coincide with the divergent term in the monolayer case; 
see Appendix C for details.
To be precise, the divergences are removed, to $O(V_{c})$ of our present interest, 
by rescaling 
\begin{equation}
v_{0}=Z\, v_{0}^{\rm ren},\ \gamma_{1}=Z\, \gamma_{1}^{\rm ren},
\label{RenormTr}
\end{equation}
with a common factor $Z$.

This scaling tells us how to carry out renormalization 
in the presence of a magnetic field $B$.
Let us write,  as in Eq.~(\ref{CRintraband}) of the monolayer case, 
the excitation energy for the $L_{n} \rightarrow L_{k}$ transition
in the form
\begin{equation}
\epsilon^{k\leftarrow n}_{{\bf k}=0}
= (s_{k}\eta_{k} -s_{n}\eta_{n})\, \big(\sqrt{2}\, v_{0}/\ell 
+ c^{kn}\, V_{c}),
\label{Eknrenorm}
\end{equation}
with $\eta_{n}= \eta_{n}(1/w)$.
Note first that $v_{0}/\gamma_{1}=v_{0}^{\rm ren}/\gamma_{1}^{\rm ren}$
is invariant under renormalization; it is therefore finite 
and does not run with $B$.
Similarly, $w=(\omega_{c}/\gamma_{1})^{2} 
\propto (v_{0}^{\rm ren}/\gamma_{1}^{\rm ren})^{2}B$
is invariant, and is linear in $B$.
This means that $\eta_{n}(1/w)$ remain unrenormalized and finite.
Equation~(\ref{Eknrenorm}) then reveals a remarkable 
structure of the Coulombic corrections $c^{kn}$:
The divergent pieces are common to all $c^{kn}$ and are removed 
by a single counterterm $\propto (Z-1)\, v_{0}^{\rm ren}$.

Figure~3~(b) depicts the momentum profiles
$\gamma_{\bf p}^{2}\, [\cdots]$ of 
$\triangle \epsilon^{k\leftarrow n}_{{\bf k}= 0}/ (s_{k}\eta_{k} - s_{n}\eta_{n})$ 
for some typical resonances.
For comparison the profile for the monolayer resonance
$(\triangle \epsilon^{1\leftarrow 0}_{{\bf k}=0})^{\rm mono}$
is also included there.
The gradually decreasing high momentum tails, common to all, 
numerically demonstrate the validity of the scaling~(\ref{RenormTr})
and Eq.~(\ref{Eknrenorm}).
This further verifies that the leading logarithmic velocity renormalization
is formally the same for both monolayer and bilayer graphene.

For renormalization let us 
refer to a specific resonance, e.g., the $-3\rightarrow -2$ resonance at $\nu=-8$,
and define $ v_{0}^{\rm ren}$ so as to absorb its entire $O(V_{c})$ correction,
\begin{equation}
\omega_{c}^{\rm ren}\equiv \sqrt{2}\, v_{0}^{\rm ren}|_{B}/\ell = \sqrt{2}\, v_{0}/\ell 
+ c^{-2,-3}\, V_{c}.
\end{equation}
One then has, for general $n\rightarrow k$ channels,
\begin{equation}
\epsilon^{k\leftarrow n}_{{\bf k}=0}
= (s_{k}\eta_{k} -s_{n}\eta_{n})\, \big(\omega_{c}^{\rm ren}
+ \triangle c^{kn}\, V_{c}).
\label{EkzeroBL}
\end{equation}
Here $\triangle c^{kn} \equiv c^{kn} -c^{-2,-3}$ are now free 
from the UV divergence and are uniquely fixed as genuine quantum corrections.
In terms of the bilayer cyclotron frequency 
$(\omega^{\rm bi})^{\rm ren} 
\equiv(\omega_{c}^{\rm ren})^{2}/\gamma_{1}^{\rm ren}$,
this also reads
\begin{equation}
\epsilon^{k\leftarrow n}_{{\bf k}=0}
= (s_{k}\zeta_{k} -s_{n}\zeta_{n})\, \big\{ (\omega^{\rm bi})^{\rm ren}
+ \triangle c^{kn}\,\sqrt{w}\, V_{c}\big\},
\end{equation}
with $\zeta_{n}=\sqrt{|n|(|n|-1)}\, /\xi_{n}(w)$
and $w=(\omega^{\rm bi})^{\rm ren}/\gamma_{1}^{\rm ren}$.

The quantum corrections $\triangle c^{kn}$,
unlike those of the monolayer case,
are not pure numbers and, actually, are functions of 
$\sqrt{w}= \omega_{c}^{\rm ren}/\gamma_{1}^{\rm ren}$.
This is seen if one notes that 
$g^{kn}_{\bf p}$ are functions of $\sqrt{w}$
and $\ell {\bf p}$ so that
$c^{kn}$ are functions of $\sqrt{w}$ and 
the cutoff $N_{L} \propto \Lambda^{2}/(eB)$; 
the cutoff-independent corrections $\triangle c^{kn}$ 
thus depend on $w$ alone.
Let us set $\gamma_{1}^{\rm ren}=\hat{\gamma}_{1}\times 100$ meV 
and
$v_{0}^{\rm ren}=\hat{v}_{0} \times 10^{6}$ m/s
so that
$1/\sqrt{w}=\gamma_{1}^{\rm ren}/\omega_{c}^{\rm ren} \approx 2.75\, G$ 
with $G= \hat{\gamma}_{1}/(\hat{v}_{0} \sqrt{B[{\rm T}]})$;
$G=1$ for $\hat{\gamma}_{1}=3.5$ and $\hat{v}_{0}\approx 1.107$ 
at $B=10$T.
It turns out that $\triangle c^{k,n}$, 
when plotted in $G$, 
behave almost linearly around $G=1$.

The way  $v_{0}^{\rm ren}$ runs with $B$ is determined from
\begin{equation}
v_{0}^{\rm ren}|_{B} = v_{0}^{\rm ren}|_{B_{0}} 
+ \delta c^{-2,-3}\, \alpha/(\sqrt{2}\, \epsilon_{b}),
\label{vzeroBL}
\end{equation}
where $\delta c^{-2,-3}\equiv c^{-2,-3}|_{B} -c^{-2,-3}|_{B_{0}}$.
Numerically $\delta c^{-2,-3}$ is nearly twice as large as 
the monolayer expression
$- (\sqrt{2}/8)\, \log (B/B_{0})$
over the range  $0.5< B/B_{0}< 2$ around $G=1$.
The decrease in $v_{0}^{\rm ren}|_{B}$ with $B$ is larger
in bilayer graphene and may amount to about 7\% for $B/B_{0}\sim 2$ 
(and $\epsilon_{b} \sim 5$).
In this way, the renormalized velocity 
$v_{0}^{\rm ren}$ is in general different,
in magnitude and running with $B$, 
for monolayer and bilayer graphene; 
it reflects their low-energy features as well.

We are now ready to look into some typical channels of cyclotron resonance.
We use Eq.~(\ref{EkzeroBL}) and evaluate 
$\triangle c^{k,n}\equiv \triangle c^{k\leftarrow n}$ numerically;
for the bilayer the filling factor $\nu = 4\,(n_{\rm f} +1)$ for $n_{\rm f} \le -2$
while  $\nu = 4\,n_{\rm f}$ for $n_{\rm f}\ge 1$.
For intraband channels one finds 
\begin{eqnarray}
\triangle c^{\pm1,-2}
&\stackrel{\nu=-4}{=}&   0.7270 + 0.5484\, \delta G,  \nonumber\\
\triangle c^{-2,-3}
&\stackrel{\nu=-8}{=}&   0,  \nonumber\\
\triangle c^{-3, -4}
&\stackrel{\nu=-12}{=}&   -0.1521 - 0.0453\,  \delta G,  \nonumber\\
\triangle c^{-4,-5}
&\stackrel{\nu=-16}{=}& -0.2496 -0.0797\, \delta G,
\label{BLinter}
\end{eqnarray}
where $\delta G = G-1$ with $G= \hat{\gamma}_{1}/(\hat{v}_{0} \sqrt{B[{\rm T}]})$.
Similarly, for interband resonances one obtains
\begin{eqnarray}
\triangle c^{3\leftarrow -2}
&\stackrel{\nu= -4}{=}&   0.3922 - 0.0023\, \delta G,  \nonumber\\
\triangle c^{2\leftarrow -3}
&\stackrel{\nu= -4}{=}&  0.4794+ 0.0706\, \delta G,  \nonumber\\
\triangle c^{2\leftarrow -3}
&\stackrel{\nu= -8}{=}& 0.3872 + 0.0552\, \delta G,  \nonumber\\
\triangle c^{3\leftarrow -4}
&\stackrel{\nu= -12}{=}& 0.2961 + 0.00145\, \delta G; 
\end{eqnarray}
also $\triangle c^{4\leftarrow -3} \stackrel{\nu= -4}{=} 0.41 + \cdots$
and $\triangle c^{4\leftarrow -3}
\stackrel{\nu= -8}{=} 0.29 + \cdots$.
These linearized expressions are numerically precise with errors of
less than 3\% over the range $0.3 < G<1.5$.
The many-body effect is thus expected to be sizable in bilayer graphene. 
An effective variation in $v_{0}^{\rm ren}$ would amount to 
about $0.7\, V_{c}/\omega_{c}^{\rm ren} \sim 20\%$ 
for $\epsilon^{1\leftarrow -2}_{{\bf k}=0}|_{\nu=-4}$, and
about -5\% for $\epsilon^{-3\leftarrow -4}_{{\bf k}=0}|_{\nu= -12}$, 
in comparison with $\epsilon^{-2\leftarrow -3}_{{\bf k}=0}|_{\nu=-8}$.

\begin{figure}[tbp]
\includegraphics[scale=0.35]{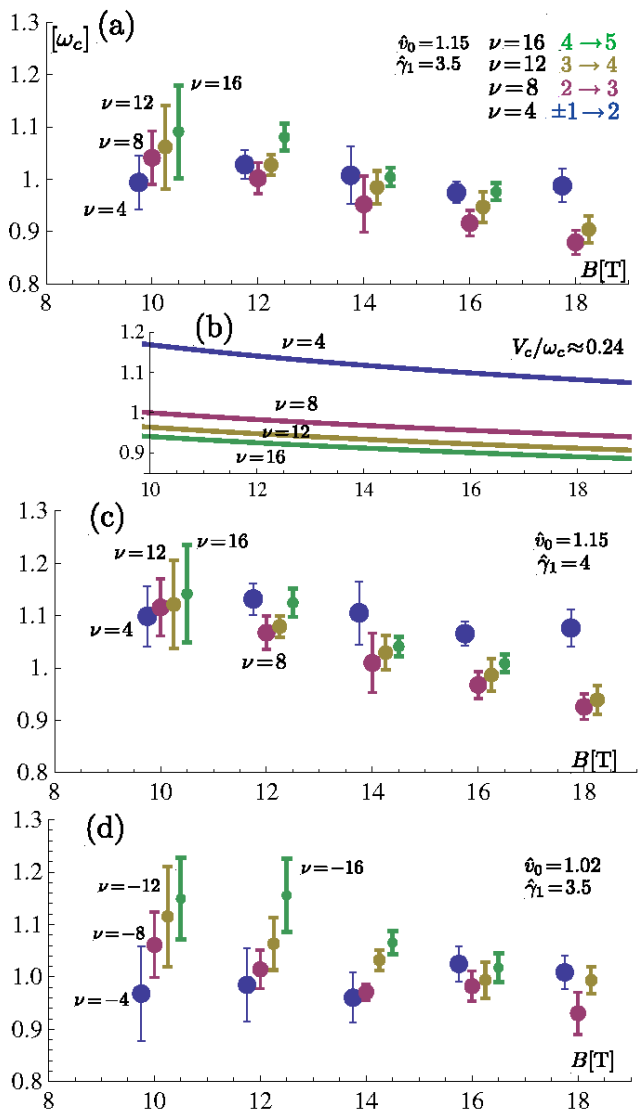}  
\caption{
Experimental data of Ref.~\onlinecite{HJTS}, reorganized in the form  
$\epsilon^{k\leftarrow n}_{{\bf k}=0}/(s_{k}\eta_{k} -s_{n}\eta_{n})$
and plotted in units of $\omega_{c}=\sqrt{2}\, v_{0}/\ell^{2}$
(with $v_{0} = 1.15 \times 10^{6}$m/s).
(a)~Electron data, analyzed with $\hat{v}_{0}=1.15$ and $\hat{\gamma}_{1}=3.5$;
for clarity the data points, originally 
at $B$=(10, 12, 14, 16)\,T,
are slightly shifted in $B$.
(b)~Theoretical expectation according to Eq.~(\ref{EkzeroBL}), with  
$\hat{v}_{0}=1.14$, $\hat{\gamma}_{1}=3.5$ 
and $V_{c}/\omega_{c}\approx 0.24$ (or $\epsilon_{b}\approx 5.6)$.
(c)~Electron data, reanalyzed with $\hat{\gamma}_{1} =4$.
(d)~Hole data, analyzed with $\hat{v}_{0} =1.02$ and $\hat{\gamma}_{1} =3.5$.
}
\end{figure}


As for experiment, Henriksen {\it et al.}~\cite{HJTS} measured, 
via IR spectroscopy, 
cyclotron resonance in bilayer graphene in magnetic fields up to 18T. 
They observed intraband transitions, which are identified with
$\{\epsilon^{2\leftarrow 1}_{{\bf k}=0}|_{\nu=4}$,
$\epsilon^{3\leftarrow 2}_{{\bf k}=0}|_{\nu=8}$,
$\epsilon^{4\leftarrow 3}_{{\bf k}=0}|_{\nu=12}$,
$\epsilon^{5\leftarrow 4}_{{\bf k}=0}|_{\nu=16}\}$ and 
the corresponding hole resonances listed in Eq.~(\ref{BLinter}),
together with an appreciable asymmetry between the electron and hole data.

Figure~3~(c) reproduces the electron data of Ref.~\onlinecite{HJTS}.
There the real curves represent the resonance energies~(\ref{EkzeroBL}) 
for $V_{c}=0$, with  $\hat{v}_{0}\approx 1.15$ deduced from the $\nu=4$ data
and $\hat{\gamma}_{1}$ taken to be 3.5, as supposed in Ref.~\onlinecite{HJTS}.
They poorly fit the $\nu=8,12,16$ data. 
Unfortunately, inclusion of the $O(V_{c})$ corrections
scarcely improves the fit, as seen from the dotted curves.

The situation becomes clearer if one, in view of Eq.~(\ref{EkzeroBL}), reorganizes
the experimental data in the form 
$\epsilon^{k\leftarrow n}_{{\bf k}=0}/(s_{k}\eta_{k} -s_{n}\eta_{n})$
and plots them in units of $\omega_{c}=\sqrt{2}\,v_{0}/\ell$
(with $v_{0} = 1.15 \times 10^{6}$m/s).
Figure~4~(a) shows such a  plot for the electron data; 
for clarity the data points for different channels, originally 
at $B$=(10, 12, 14, 16)\,T, 
are slightly shifted in $B$.
It is to be contrasted with Fig.~4~(b), which illustrates 
how each resonance would behave with $B$, according to Eq.~(\ref{EkzeroBL}), 
for $V_{c}\approx 10 \sqrt{B[{\rm T}]}$ meV (or $\epsilon_{b}\approx 5.6)$;
 in particular, the $\nu=8$ curve represents the running of $v_{0}^{\rm ren}|_{B}$ 
according to Eq.~(\ref{vzeroBL}).
In Fig.~4~(a) the $\nu=8,12,16$ resonances are apparently ordered in a way opposite 
to Fig.~4 (b), and an appreciable gap between the $\nu=4$ resonance 
and the rest is not very clear. 
The $\nu=8,12,16$ data show a general trend 
to decrease with $B$, consistent with possible running of $v_{0}^{\rm ren}|_{B}$ 
but at a rate faster than expected.
It is rather difficult to interpret these features, but they, in part, 
could be attributed to possible quantum screening~\cite{KSpzm} of 
the Coulomb interaction in bilayer graphene 
such that $\epsilon_{b}$ is effectively larger~\cite{fntwo} for lower $B$.
Note, in this connection, Fig.~4~(c) which shows that the same data may 
suggest a Coulombic gap for a choice $\gamma_{1}\approx 4$
favored in Ref.~\onlinecite{ZLBF}.
We further remark that, in spite of an asymmetry in electron and hole data, 
the hole data shares essentially the same features;
see Fig.~4~(d).

No data are available for interband cyclotron resonance in bilayer graphene at present.
They are highly desired 
because the comparison of interband and intraband resonances 
from the same initial states
would provide a clearer signal for the many-body effect.
We record the ratios 
\begin{eqnarray}
\triangle R(-3 \rightarrow \pm 2) &\stackrel{\nu=-8}{\approx}&
0.39\, V_{c}/\omega_{c}^{\rm ren} \sim 15\%,
\nonumber\\
\triangle R(-4 \rightarrow \pm 3)&\stackrel{\nu=-12}{\approx}&
0.45\, V_{c}/\omega_{c}^{\rm ren} \sim 18\%,
\label{ratioBL}
\end{eqnarray}
which imply that a close look into 
the $(-3 \rightarrow  \pm 2)$ resonances 
and the $(-4 \rightarrow  \pm 3)$ resonances 
would find a sizable variation $\sim 15 \%$  in $v_{0}^{\rm ren}$.

\section{Summary and discussion}

Graphene supports charge carriers that behave as Dirac fermions,
which,  in a magnetic field, lead to a characteristic
particle-hole symmetric pattern of Landau levels.
Accordingly, unlike standard QH systems, there is a rich variety 
of cyclotron resonance, both intraband and interband resonances 
of various energies, in graphene.

In this paper we have studied many-body corrections to 
cyclotron resonance in graphene.
We have constructed an effective theory using the SMA and noted that
genuine nonzero many-body corrections 
(not due to fine splitting in spin or valley)  derive from 
the quantum fluctuations of the vacuum (the Dirac sea).
Such quantum corrections are intrinsically ultraviolet divergent 
and, as we have emphasized, it is necessary to carry out renormalization of 
velocity $v_{0}$  
(and, for bilayer graphene, interlayer coupling $\gamma_{1}$ as well)
to determine the many-body corrections uniquely in terms of physical quantities.
As a result, the observable intralayer and interlayer coupling strengths 
$v^{\rm ren}_{0} \propto \gamma^{\rm ren}_{0}$ 
and $\gamma^{\rm ren }_{1}$ 
in general run with the magnetic field $B$.

Experimental data on cyclotron resonance generally have sizable error bars,
which make a clear identification of the many-body effect difficult.  
In this respect, we have presented a way to analyze the data, 
as in  Fig.~2~(b) and Fig.~4,
with the effect of renormalization properly taken into account.

For monolayer graphene a piece of data~\cite{JHT} 
which compares some leading  interband and intraband resonances 
is apparently consistent 
with the presence of many-body corrections
roughly in magnitude and sign, and also in the running of 
$v_{0}^{\rm ren}$ with $B$.

For bilayer graphene the existing data are only for intraband resonances 
and are rather puzzling, as discussed in Sec.~IV.
They generally appear to defy good fit by theory but certainly suggest 
nontrivial features of many-body corrections, such as running with $B$.

More precise measurements of cyclotron resonances are highly desired. 
Of particular interest are experiments which compare 
interband and intraband resonances from the same initial states,
as listed in Eqs.~(\ref{ratioML}) and (\ref{ratioBL}), which would clarify 
the many-body effect with minimal uncertainties.

\acknowledgments

This work was supported in part by a Grant-in-Aid for Scientific Research
from the Ministry of Education, Science, Sports and Culture of Japan 
(Grant No. 21540265).

\appendix

\section{Landau levels in bilayer graphene}

This appendix summarizes the effective Hamiltonian and 
its eigenfunctions for bilayer graphene in a magnetic field $B$. 
The bilayer Hamiltonian with interlayer coupling 
$\gamma_{1}\equiv \gamma_{A'B}$ 
is written, at one ($K$) valley, as~\cite{MF} 
\begin{equation}
H^{\rm bi} = \left(
\begin{array}{cccc}
0  & &  &  v_{0}\,\Pi^{\dag} \\
  &0 &  v_{0}\,\Pi &  \\
 &  v_{0}\,  \Pi^{\dag}  & 0 & \gamma_{1} \\
v_{0} \Pi & &  \gamma_{1}  & 0 \\
\end{array}
\right),
\label{Hbi}
\end{equation}
which acts on an electron field of the form
$\Psi_{K} = (\psi_{A},\psi_{B'},\psi_{A'}, \psi_{B})^{\rm t}$ 
in obvious notation;
$\Pi = \Pi_{x} - i\Pi_{y}$ and $\Pi^{\dag} = \Pi_{x} + i\Pi_{y}$,
with $A_{i}\rightarrow B(-y,0)$.

The energy eigenvalues obey the equation 
\begin{equation}
(|n|-1 - {\epsilon'}^{2})\, 
(|n|  -  {\epsilon'}^{2}) - {\gamma'}^{2}\, (\epsilon')^{2} =0,
\end{equation}
where $\epsilon'\equiv \epsilon_{n}/\omega_{c}$,
$\gamma' \equiv  \gamma_{1}/\omega_{c}$
and $\omega_{c}=\sqrt{2}\, v_{0}/\ell$.
This leads to the two branches of spectra $(\epsilon_{n}, \epsilon^{+}_{n})$ 
in Eqs.~(\ref{EnBL}) and ~(\ref{highenegyBr}).
In particular, zero energy $\epsilon_{n}=0$ is possible for $|n|=0$ or $|n|=1$
while  $|\epsilon^{+}_{\pm1}| > \gamma_{1}$.
A weak interlayer voltage ${1\over{2}}(\delta V)\, {\rm diag}[1,-1,-1,1]$,
added to $H^{\rm bi}$, reveals that the zero modes 
actually have $n=0_{+}$ and $n=1$ for $\delta V>0$.

The corresponding eigenfunctions for $n=\pm 2, \pm 3,...$ 
take the form
\begin{eqnarray}
\Psi_{n} &=& D_{n}\left(
\begin{array}{l}
|\, |n|\, \rangle\\ 
\beta_{n}^{(2)}\, ||n|-2\rangle\\
\beta_{n}^{(3)}\, ||n|-1\rangle\\
\beta_{n}^{(4)}\, ||n|-1\rangle\\
\end{array}\right),
\label{Psin} \\
\beta_{n}^{(2)}&=&{\sqrt{|n|-1}\over{\epsilon'}}\, \beta_{n}^{(3)},\ 
\beta_{n}^{(3)}= -{1\over{\gamma'}}\, {|n|-{\epsilon'}^{2}\over{\sqrt{|n|}}},\  
\nonumber\\
\beta_{n}^{(4)}&=& {\epsilon'\over{\sqrt{|n|}}}, 
D_{n} = {1\over{\sqrt{2}}}\, 
\sqrt{|n|(|n|-1 -{\epsilon'}^{2})\over{ |n|(|n|-1) -{\epsilon'}^{4}}},
\end{eqnarray}
where only the orbital eigenmodes are shown 
using the standard harmonic-oscillator basis $\{|n\rangle\}$. 
These expressions for $\Psi_{n}$ are equally valid 
for both the low- and high-energy branches $\epsilon_{n}$ and $\epsilon^{+}_{n}$ 
of Landau levels, depending on $\epsilon'$ one employs.

The zero-energy eigenmodes are given by 
\begin{eqnarray}
&&\Psi_{0_{+}} \! =\,\, (|0\rangle, 0,0,0)^{t},\nonumber\\
&&\Psi_{1} = D_{1}\,\big(|1\rangle, 0,-(1/\gamma')\, |0\rangle,0\big)^{t},
\end{eqnarray}
with $D_{1}=(1+ 1/{\gamma'}^{2})^{-1/2}$.

At another ($K'$) valley the Hamiltonian is given by Eq.~(\ref{Hbi})  with 
$v_{0}\rightarrow -v_{0}$ and acts on a field of the form 
$\Psi_{K'} = (\psi_{B'},\psi_{A},\psi_{B}, \psi_{A'})^{\rm t}$.
Accordingly one finds that
\begin{eqnarray}
\epsilon_{n}|_{K'} &=& - \epsilon_{-n}|_{K}, 
D_{n}|_{K'} = D_{-n}|_{K},
\beta_{n}^{(2)}|_{K'} = \beta_{-n}^{(2)}|_{K},\nonumber\\
\beta_{n}^{(3)}|_{K'}&=& - \beta_{-n}^{(3)}|_{K},
\beta_{n}^{(4)}|_{K'} = - \beta_{-n}^{(4)}|_{K}.
\end{eqnarray}
The zero-energy levels now have $n=0_{-}$ and $n=-1$.

\section{Coupling to current}

Consider a weak time-varying vector potential $(A_{x}(t), A_{y}(t))$ coupled 
to the total current in graphene.  For the effective Lagrangian $L_{\Phi}$
in Eq.~(\ref{Lphi}) 
this yields  coupling of $A_{i}$ to $\Phi^{kn}_{{\bf k}=0}
= \int d^{2}{\bf x}\, \Phi^{kn}$ of the form
\begin{eqnarray}
H_{A}
&=& -i{e\ell \omega_{c}\over{\sqrt{2}}} \sqrt{\rho_{0}}\, d_{n}\, 
\big\{ A\, {\Phi}^{\pm (n-1), -n}_{{\bf k}= 0}
+ A^{\dag}\,\Phi^{n, \pm (n-1)}_{{\bf k}= 0} \big\}\nonumber\\
&& + {\rm h.c.},
\label{HbiA}
\end{eqnarray}
where $d_{n}=\pm 1/2$ for $n\ge 2$ and $d_{1}= \pm 1/\sqrt{2}$;
$A=A_{x} - i A_{y}$; $\rho_{0}= 1/(2\pi \ell^{2})$.
The cyclotron resonance thus obeys the selection rule $\triangle |n|=\pm 1$.
In particular, the $-n \rightarrow \pm (n-1)$ transitions and 
the $\pm (n-1)\rightarrow n$ transitions $(n\ge 1)$ are distinguished~\cite{AFAC} 
by use of circularly-polarized light $\propto A_{x} \pm i A_{y}$.

Equation~(\ref{HbiA}) (with $n\ge 2)$ applies to the case of bilayer graphene as well
if one sets  $\omega_{c}\rightarrow \omega_{c}^{\rm bi}$,
$A^{\dag}\rightarrow -A^{\dag}$, 
$d_{n}= \sqrt{n-1}$ for $n\ge 3$ and $d_{2}= \sqrt{2}$,
apart from terms of $O( (\omega_{c}/\gamma_{1})^2 )$.

\section{propagators}

In this appendix we derive the electron propagator 
for bilayer graphene in free space.  
Let us set $\Pi \rightarrow p_{x}- ip_{y}$ in $H^{\rm bi}$ of Eq.~(\ref{Hbi}) 
and consider the propagator
\begin{equation}
\langle \Psi (x) \Psi^{\dag}(x') \rangle
= \langle x| 1/(i\partial_{t} -H^{\rm bi}) |x' \rangle
\end{equation}
with $|x\rangle \equiv|{\bf x}\rangle\, | t \rangle$.
We divide the $4\times 4$ matrix $H^{\rm bi}$ into a $2\times 2$ block
form and invert $(i\partial_{t} -H^{\rm bi})$.
In Fourier $({\bf p},\omega)$ space the propagator reads,
in $2\times 2$ block form, 
\begin{eqnarray}
\langle \Psi \Psi^{\dag}\rangle_{11} &=&\!
{i\over{D}}\, \Big\{\omega\, ( \omega^{2}\! - v_{0}^{2}\, {\bf p}^{2}\! 
-\gamma_{1}^{2}) + \gamma_{1} v_{0}^{2} P \sigma_{1} P \Big\},
\nonumber\\
\langle \Psi \Psi^{\dag}\rangle_{21} 
&=& {i\over{D}}\, (\omega^{2}  - v_{0}^{2}\, {\bf p}^{2} 
+\omega\, \gamma_{1}\, \sigma_{1})\,  v_{0} P, \nonumber\\
\langle \Psi \Psi^{\dag}\rangle_{12}  &=& {i\over{D}}\, v_{0} P\, 
(\omega^{2} - v_{0}^{2}\, {\bf p}^{2}   +\omega\,  \gamma_{1}\, \sigma_{1}), 
\nonumber\\
\langle \Psi \Psi^{\dag}\rangle_{22} 
&=&  i\, {\omega \over{D}}\, (\omega^{2} -v_{0}^{2}\, {\bf p}^{2}\,  
+\omega\,  \gamma_{1}\, \sigma_{1}),
\end{eqnarray}
where $D= (v_{0}^{2}\, {\bf p}^{2} -\omega^{2})^{2} 
- \omega^{2}  \gamma_{1}^{2}
= (\omega^{2}- E_{+}^{2})(\omega^{2}- E_{-}^{2})$
with $E_{\pm}= \sqrt{\gamma_{1}^{2}/4+\! v_{0}^{2}\, {\bf p}^{2}} 
\pm \gamma_{1}/2$;
$P= p^{\dag}\, \sigma_{+}
+p\, \sigma_{-}$ and $P\sigma_{1}P = (p^{\dag})^{2}\, \sigma_{+}
+p^{2}\, \sigma_{-}$ with 
 $p=p_{x}-ip_{y}$, $p^{\dag}=p_{x}+ip_{y}$ and 
$\sigma_{\pm}= (\sigma_{1}\pm i \sigma_{2})/2$.

This leads to the instantaneous propagator 
$\int (d \omega/2\pi)\langle \Psi \Psi^{\dag}\rangle$,
\begin{equation}
\langle \Psi \Psi^{\dag}\rangle|_{t=t'} 
= {1\over{4{\cal D}_{\bf p}}}\left(
\begin{array}{cc}
\gamma_{1}\, P \sigma_{1} P/{\bf p}^{2} &  2v_{0}\,  P\\
2v_{0}\,  P & \gamma_{1}\, \sigma_{1} \\
\end{array}
\right),
\end{equation}
with ${\cal D}_{\bf p}=\sqrt{\gamma_{1}^{2}/4 + v_{0}^{2}\, {\bf p}^{2}}$.

To calculate the Coulomb exchange correction one may 
replace, in Eq.~(\ref{vS}), $iS({\bf p})$ by this propagator. 
Note that $v_{0}\, P/(2\,{\cal D}_{\bf p})$ approaches, 
for ${\bf p}\rightarrow \infty$, the monolayer propagator 
$iS({\bf p}) = \sigma_{i}p_{i} /(2\, |{\bf p}|)$
(apart from an inessential mismatch 
$\sigma_{2} \rightarrow -\sigma_{2}$ in notation).
As a result, setting $iS({\bf p}) \rightarrow v_{0}\, P/(2\,{\cal D}_{\bf p})$
for $v_{0}$
and  $iS({\bf p}) \rightarrow \gamma_{1}/(4\,{\cal D}_{\bf p})$
for $\gamma_{1}$ and carrying out the ${\bf k}$ integration, 
as in Eq.~(\ref{vS}), yield 
the same amount of logarithmic divergence
$\sim (\alpha/8 \epsilon_{b})\,  \log \Lambda^{2}$ 
as in the monolayer case;
it thus renormalizes $v_{0}$ and $\gamma_{1}$ simultaneously 
as in Eq.~(\ref{RenormTr}).



\begin{thebibliography}{99}
\bibitem{fnzero}
The need for velocity and mass renormalization for electrons in bilayer graphene 
was earlier discussed in Ref.\onlinecite{KCC}, 
with a phenomenological fit to the quasiparticle dispersion 
within the Thomas-Fermi approximation and Hartree-Fock theory.
\bibitem{fnone} 
Here ${\cal C}_{N}= (\sqrt{2}/8)\sum_{k=1}^{N_{L}}\Gamma (k-1/2)\, \sqrt{k}/k!$,
which is related to the quantities $a$ in Ref.~\onlinecite{IWFB} and $C_{1}$  
in Ref.~\onlinecite{BMgr} as
${\cal C}_{N}= a/V_{c} - (3/4)\, \sqrt{\pi/2} = C_{1} - (3/8)\, \sqrt{\pi/2}$.
\end{thebibliography}
\end{document}